\documentclass[prb,preprint,a4paper,superscriptaddress,showpacs]{revtex4}

\usepackage{bm}
\usepackage[dvips]{graphicx}
\usepackage{hyperref}
\usepackage{amsmath}
\usepackage{enumitem}
\usepackage{color}

\begin{document}

\author{Ryo KOBAYASHI}
\email{kobayashi.ryo@nitech.ac.jp}
\affiliation{Department of Scientific and Engineering Simulation, Nagoya Institute of Technology, Gokiso-cho, Showa-ku, Nagoya 466-8555, Japan}
\affiliation{Center for Materials research by Information Integration, National Institute for Materials Science, 1-2-1 Sengen, Tsukuba 305-0047, Japan}
\author{Tomoyuki TAMURA}
\affiliation{Department of Scientific and Engineering Simulation, Nagoya Institute of Technology, Gokiso-cho, Showa-ku, Nagoya 466-8555, Japan}
\affiliation{Center for Materials research by Information Integration, National Institute for Materials Science, 1-2-1 Sengen, Tsukuba 305-0047, Japan}
\author{Ichiro TAKEUCHI}
\affiliation{Department of Scientific and Engineering Simulation, Nagoya Institute of Technology, Gokiso-cho, Showa-ku, Nagoya 466-8555, Japan}
\affiliation{Center for Materials research by Information Integration, National Institute for Materials Science, 1-2-1 Sengen, Tsukuba 305-0047, Japan}
\author{Shuji OGATA}
\affiliation{Department of Scientific and Engineering Simulation, Nagoya Institute of Technology, Gokiso-cho, Showa-ku, Nagoya 466-8555, Japan}

\title{Sparse selection of bases in neural-network potential for crystalline and liquid Si}

\begin{abstract}
 The neural-network interatomic potential for crystalline and liquid Si has been developed 
 using the forward stepwise regression technique to reduce the number of bases
 with keeping the accuracy of the potential.
 This approach of making the neural-network potential enables us to construct the accurate interatomic potentials 
 with less and important bases selected systematically and less heuristically.
 The evaluation of bulk crystalline properties, and dynamic properties of liquid Si
 show good agreements between the neural-network potential and {\itshape ab-initio} results.
\end{abstract}



\maketitle

 \section{Introduction}\label{sec:intro}

 Molecular dynamics (MD) is a powerful simulation tool for atomistic- to nano-scale dynamic phenomena
 and nowadays widely used in a variety of research field such as crystal growths,\cite{Matsumoto:2002km} 
 defect formations,\cite{Henriksson:2006ig,Kobayashi:2015hr} ion bombardment,\cite{Greaves:2013dn}
 radiation damages,\cite{Kobayashi:2015hr} ion conduction,\cite{Jalem:2013fn} thermal conduction,\cite{Tanaka:2015kj}
 and chemical reactions.\cite{Otani:2006hz,Heyden:2005kt}
 The validity of those simulation results depends strongly on the interatomic potentials used.
 An {\itshape ab-initio} MD is an extreme example of the accurate interatomic potential,
 however, since it costs significantly more than the empirical potentials, 
 it is not suitable for large-scale or long-time simulation.
 Thus a lot of empirical interatomic potentials have been proposed since the MD was invented.
 However, most empirical potentials developed so far reproduce limited physical quantities of
 a certain situation, since they have only a few basis functions such as one or two two-body terms, a three-body term,
 and a functional term.\cite{Carlsson:1990tl}

 Under this circumstance, some studies of making more accurate
 interatomic potentials applicable for a variety of situations 
 using many basis functions and machine-learning techniques have emerged
 such as the neural network (NN) potential,\cite{Behler:2007fe,Behler:2008ft,Behler:2008cy,Behler:2011it,Artrith:2011em,Artrith:2012el,Sosso:2012hi,Artrith:2012fw,Morawietz:2013cc}
 the Gaussian approximation potential (GAP),\cite{Bartok:2010fj}
 and the linear regression (LR) potential.\cite{Seko:2014ke}
 Since these potentials use more basis functions than former empirical potentials that consist of one or a few functions,
 they reproduce reference values better and usually they are more accurate than the former potentials
 whereas they are much slower than the former potentials.
 Comparing the NN and LR potentials, 
 the NN potential is expected to be more accurate than the LR potential in case of the same number of bases,
 since it has non-linear activation functions in the NN structure.
 This non-linearity is expected to work like bond-order\cite{Albe:2002gj} or environment-dependent potentials.\cite{Bazant:1998ut}
 Because of this, and also because we can control the accuracy and computational cost 
 by choosing the number of layers and nodes in a layer,
 we adopt the NN potential in this paper.
 However, it is still heuristic that what kind of and how many basis functions should be used for the NN,
 and we have to find appropriate basis functions by trial-and-error.

 Seko {\itshape et al.}\cite{Seko:2014ke} showed that the number of bases in the LR potential can be drastically reduced
 retaining its accuracy by using the least absolute shrinkage and selection operator (LASSO) technique.
 The LASSO adds a penalty term to the function to be minimized and enables us to select important bases by eliminating other bases. 
 This technique is very powerful to reduce and select bases relevant to the system considered,
 however, to the best of our knowledge, such a basis selection has not been employed in the NN potential.
 We adopt a similar sparsification technique, called the forward stepwise (FS) regression,\cite{Trevor:2001tz}
 which can usually sparsify more aggressively or reduce more bases than the LASSO.
 With using the FS regression, important bases are selected and the number of bases in NN potential is
 reduced systematically and less heuristically.

 In this paper, we create a NN potential for crystalline and liquid Si system,
 since in spite of a lot of efforts to make accurate and general-purpose interatomic potential for Si system,
 there is no sufficient one that can be used over the wide range of situation 
 and well reproduces physical properties of crystalline structures, amorphous structures, and dynamic properties of liquid.
 For example, the Stillinger-Weber (SW) potential does not well reproduce some physical properties of a crystalline structure
 and needs further refinement of parameters,\cite{Stillinger:1985co,Pizzagalli:2013dr}
 and the power spectra of velocity auto-correlation calculated using SW potential and Tersoff potential\cite{Tersoff:1988to,Tersoff:1988vk}
 do not agree with that of an {\itshape ab-initio} calculation.\cite{Ishimaru:1996us,Stich:1996cv}
 This paper compares some physical properties of a variety of situations obtained with the NN potential 
 to those obtained with other potential and the density-functional theory (DFT) calculation in the Sec.~\ref{sec:results}.


 \section{Computation details}\label{sec:comp}

\begin{figure}
   \begin{center}
    \includegraphics[keepaspectratio,width=.4\textwidth]{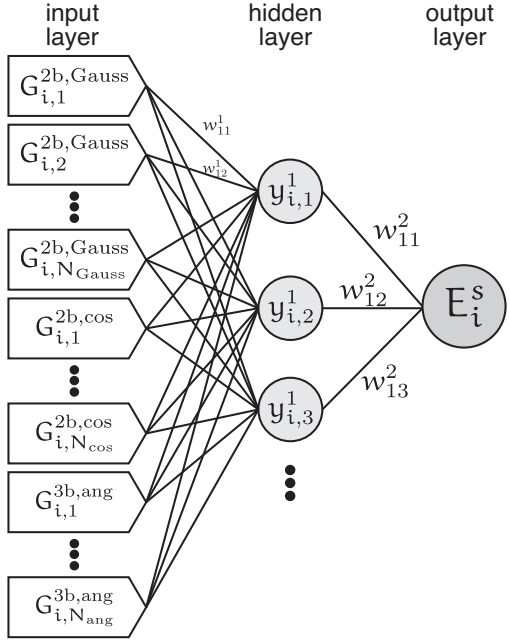}
    \caption{The structure of 1-hidden-layer NN model for Si system.}
    \label{fig:NN1-structure}
   \end{center}
  \end{figure}

  We adopt the 1-hidden-layer NN model originally developed by Behler and Parrinello\cite{Behler:2007fe}
  as shown in Fig.~\ref{fig:NN1-structure}.
  The energy of an atom-$i$ in a structure-$s$ is defined as the following equations,
  \begin{eqnarray}
   E_i^s &=& \sum_l w^2_{1l} y^1_{i,l},  \label{eq:E_i}\\
   y^1_{i,m} &=& \sigma \left( \sum_n w^1_{mn} G_{i,n}^{\rm type}\right).\label{eq:y1}
  \end{eqnarray}
  Here $w^l_{mn}$ is the weight of the line from the node-$n$ in $(l-1)$-th layer to the node-$m$ in $l$-th layer,
  $y^l_{i,m}$ is the value of node-$m$ in $l$-th layer,
  and $G_{i,n}^{\rm type}$ is the $n$-th symmetry function of a certain type, which depends on interatomic bond distances from the atom-$i$ 
  or angles between bonds around atom-$i$.
  Superscripts on $G_{i,n}^{\rm type}$ indicate the type of symmetry function that can be 2-body or 3-body function.
  The activation function $\sigma(x)$ is defined using the sigmoid function as,
  \begin{equation}
   \sigma(x) = \frac{1}{1+ {\rm e}^{-x}} -\frac{1}{2}.
  \end{equation}
  Here we subtract 1/2 from the sigmoid function
  since the activation function is required to be zero when inputs are zero.

  The symmetry functions include two types for 2-body functions, Gaussian and cosine, and
  one 3-body angular function:
  \begin{eqnarray}
   G^{\rm 2b,Gauss}_{i,n} & \equiv & \sum_{j\neq i} {\rm e}^{-\eta_n r_{ij}^2} f_{\rm c}(r_{ij}), \label{eq:Gauss}\\
   G^{\rm 2b,cos}_{i,n} & \equiv & \sum_{j \neq i} (1+\cos \xi_n r_{ij}) f_{\rm c}(r_{ij}), \label{eq:cos}\\
   G^{\rm 3b,ang}_{i,n} & \equiv & \sum_{j,k \neq i} \left( \lambda_n +\cos \theta_{ijk} \right)^2 f_{\rm c}(r_{ij}) f_{\rm c}(r_{ik}), \label{eq:ang}
  \end{eqnarray}
  where $\eta_n$, $\xi_n$, $\lambda_n$ are the parameters, $r_{ij}$ is the interatomic distance between $i$ and $j$,
  $\theta_{ijk}$ is the angle between bonds $i$-$j$ and $i$-$k$,
  and the cutoff function  $f_{\rm c}(r)$  is defined as
  \begin{equation}
   f_{\rm c}(r)=
    \left\{ \begin{array}{ll}
     \frac{1}{2}\left[ \cos \frac{\pi r}{r_c} +1\right], &  \text{for}\ \ r \le r_{\rm c}, \\
             0, & \text{for}\ \ r > r_{\rm c}.
            \end{array}\right.
  \end{equation}
  Here $r_{\rm c}$ is the cutoff radius.
  As shown in Fig.~\ref{fig:NN1-structure}, each function type has some different parameters,
  for example, in case of the Gaussian type, there are $N_{\rm Gauss}$ parameters $\{\eta_n\}$.

  The weights in the NN potential  $\{w\}$  are optimized through minimizing the following function,
  \begin{eqnarray}
   \mathcal{L}(\{w\}) &=&  \frac{1}{N_s} \sum_s
    \left[
     \left( \frac{E^{s,{\rm NN}} -E^{s,{\rm DFT}}}{N^s_{\rm a}} \right)^2 \right. \nonumber \\
   &+& \left.\frac{\kappa}{3N^s_{\rm a}}\sum_i^{N^s_{\rm a}} \sum_{\alpha}^{x,y,z}\left( F_{i,\alpha}^{s,{\rm NN}}-F_{i,\alpha}^{s,{\rm DFT}}\right)^2
       \right],
   \label{eq:L}
  \end{eqnarray}
  where $E^{s,{\rm DFT}}$ and $E^{s,{\rm NN}}=\sum_i E^{s,{\rm NN}}_i$ is the cohesive energies of a structure-$s$ obtained 
  via DFT and NN-potential calculations;
  $F^{\rm s,DFT}_{i,\alpha}$ and $F_{i,\alpha}^{s,{\rm NN}}$ are the $\alpha$-components of forces on atom-$i$ in the structure-$s$
  obtained via DFT and NN-potential calculations.
  $N^s_{\rm a}$ is the number of atoms in a structure-$s$, and $N_s$ is the number of samples considered.
  The parameter $\kappa$ is the conversion factor for forces that absorbs the difference of units between energy (eV) and force (eV/\AA );
  it is set 0.05, which makes the force value 0.2 eV/{\AA} correspond to the energy value 0.01 eV.
  The force term is important if the potential is applied to some dynamics simulation.\cite{Ercolessi:1994vu}

  In order to select relevant and small number of bases,
  we adopt the FS regression approach.\cite{Trevor:2001tz}
  The FS regression in this study proceeds as follows:
  \begin{enumerate} 
   \item Weights on lines connecting bases and 1st layer nodes $\{w^1\}$ are set zero and other weights are set non-zero.
         Hence the weights to be optimized, $\{w\}_{\rm o}$, include only $\{w^2\}$ at this point.
   \item Compute the sum of squared gradients associated to $\{w^1\}$,
         \begin{equation}
          g^{\rm type}_n \equiv \sum_m \left(\frac{\partial \mathcal{L}}{\partial w^1_{mn\in {\rm type}}}\right)^2,
           \label{eq:g_n}
         \end{equation}
         and find the largest one $g^{\rm max}_n$.
   \item Append the weights, $\{ w^1_{mn\in {\rm max}}\}$, which is related to the basis just selected in the previous step, to 
         the weight set to be optimized, $\{w\}_{\rm o}$.
   \item Perform the minimization with respect to the current weight set $\{w\}_{\rm o}$ until
         the convergence criterion, $|\mathcal{L}^m -\mathcal{L}^{m-1}|/|\mathcal{L}^{m-1}| < \varepsilon_{\rm tol}$, is achieved
         or the iteration $m$ exceeds the maximum iteration number $M$, 
         where $\mathcal{L}^m$ is the $\mathcal{L}$ value at $m$-th iteration.
   \item Go back to 2 unless all the bases are selected.
  \end{enumerate}
  The number of iterations $M$ in the minimization at the step 4 should be large enough 
  to find the minimum or at least the basin of the minimum in the $\{w\}_{\rm o}$-space.
  However, since large $M$ results in a lot of computation time for the iteration of the outer loop,
  $M$ or the convergence criterion for the minimization is usually set moderate value determined empirically.
  Since the FS regression searches the minimum near 
  the origin of parameter space $\{w^1\}$ starting from the origin $w^1_{mn}= 0$,
  over-fitting can be suppressed and a sparse NN model can be obtained as same as the LASSO.
  The FS regression has an advantage against the LASSO with regard to the computational cost of minimization,
  since the FS does not need to evaluate gradients of all weights but only the weights to be optimized $\{w\}_{\rm o}$
  during the inner minimization loop,
  whereas the LASSO needs the gradients of all the weights.
  We can select the important bases that appears earlier than the other ones.
  
  
  The sample structures prepared in this work were chosen so that the NN potential can well reproduce 
  the crystalline and amorphous Si structures, some defect formation energies, and dynamical properties of liquid Si
  The samples include:
  perfect crystalline Si of some different crystal structure, such as diamond, fcc, bcc, and 
  hcp; crystalline structure which involves a Si vacancy; generalized stacking faults on (111) plane of diamond structure; 
  and uniform deformation, shear deformation, and random atom displacements of above mentioned structures;
  and liquid Si structures extracted from {\itshape ab-initio} MD simulation runs.
  We prepare 3930 samples that include 3930 energies and 90096 force components.
  The samples are divided into training-set data and test-set data, and the training-set data are used to optimize the NN
  and the test-set data are used to evaluate the NN potential.
  In this paper, randomly selected 10 \% of all the samples are chosen as test samples.

  To obtain the reference values, we adopt the DFT calculation
  using {\scshape vasp} (Vienna Ab-initio Simulation Package).\cite{Kresse:1996vf,Kresse:1999wc}
  The generalized gradient approximation (GGA) functional\cite{Perdew:1996iq} for the exchange-correlation potential and 
  the projector augmented wave (PAW) method\cite{Blochl:1994dx} for the pseudo-potential of Si are adopted.
  The energy cutoff of expansion of plane waves is 250 eV.
  Phonon dispersion relations are calculated using {\scshape phonopy}.\cite{Togo:2008jt}
  Molecular statics and MD calculations of liquid Si with NN potential are performed using our own code,
  {\scshape nap} (Nagoya Atomistic simulation Package).\cite{NAP:2014vx}

 \section{Results and discussion}\label{sec:results}
  
  \begin{figure}[tbp]
   \begin{center}
    \includegraphics[keepaspectratio,width=.45\textwidth]{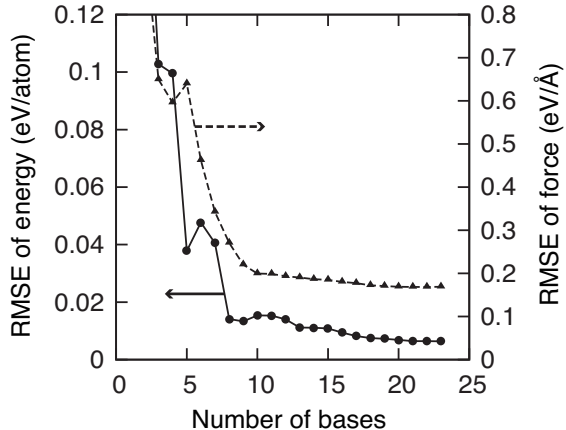}
    \caption{RMSEs of energy and force of test set as functions of number of bases extracted through the FS regression.}
    \label{fig:rmse-bases}
   \end{center}
  \end{figure}
  
\begin{figure}[tbp]
   \begin{center}
    \includegraphics[keepaspectratio,width=.4\textwidth]{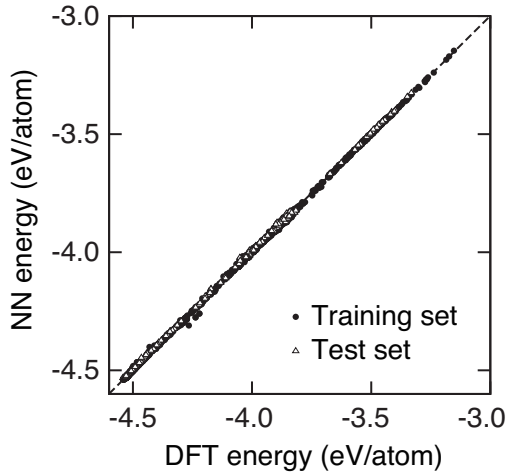}
    \caption{Relation between NN energies and DFT energies of sample data set.
    Energies about training set (filled circle) and test set (open triangle) are mostly on the ideal line.}
    \label{fig:energy-NN-DFT}
   \end{center}
  \end{figure}

  Relevant bases for crystalline and liquid Si system are selected using the FS regression as follows.
  The cutoff radius $r_{\rm c}$ and the number of nodes in a hidden layer are set 4.0 \AA ~ and 10, respectively,
  according to the previous studies\cite{Behler:2007fe,Behler:2011it} and preliminary calculations.
  We prepare 300 bases including Gaussian, cosine, and angular types as $(N_{\rm Gauss}, N_{\rm cos}, N_{\rm ang})=(100,100,100)$.
  The number of iterations in the inner minimization, $M$, at the step 4 of the FS regression is 200.
  During the FS regression, the root mean squared errors (RMSEs) of energy and force for test set decrease 
  as the number of bases increases as shown in Fig.~\ref{fig:rmse-bases}.
  Eighteen bases are chosen since the RMSE of energy is less than 10 meV/atom.
  The selected bases include four Gaussian bases with $\{\eta\}=$
  \{0.1796, 0.4184, 1.0551, 1.9306\}[${\rm \AA^{-2}}$],
  eleven cosine bases with $\{\xi\}=$ \{0.7854, 0.8816, 1.1701, 1.4105,
  1.7471, 2.4684, 2.6607, 2.8050, 2.8531, 2.9493, 3.1416\}[${\rm \AA^{-1}}$],
  and three angular bases with $\{\lambda\}=$ \{0.0, 0.6327, 1.0\}.
  Then, the weights $\{w\}$ of these 18 bases are fully optimized within the space of selected bases.
  Final values of RMSE of energy and force obtained via the minimization are 3.9 meV/atom and 173 meV/\AA , respectively.
  Figure \ref{fig:energy-NN-DFT} shows the relation between energies of samples obtained using NN potential and DFT.
  We can see that the NN potential with reduced number of bases well reproduces the DFT energies over
  a wide variety of structures.
  
\begin{figure}[tbp]
   \begin{center}
    \includegraphics[keepaspectratio,width=.4\textwidth]{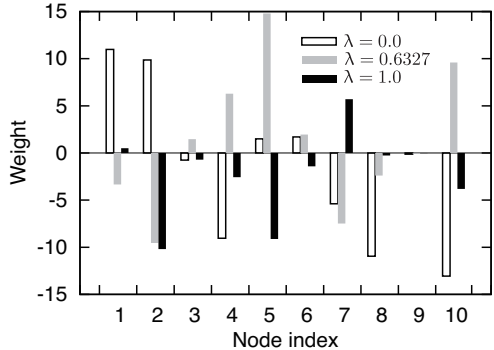}
    \caption{Weights from angular terms to nodes in the hidden layer, $\{w^1_{mn \in {\rm ang}}\}$.
    White, gray, and black bins indicate the weights from the angular bases with $\lambda=0.0$, 0.6327, and 1.0, respectively.}
    \label{fig:angular-weights}
   \end{center}
  \end{figure}

  It is worth looking into the NN structure in detail to understand how the NN potential works with selected bases.
  There are three angular bases selected through the FS regression. 
  However, the $\lambda=1/3$, which corresponds to the bond angle of diamond Si and is adopted in the SW potential,
  is not included in the selected bases.
  The angular basis with $\lambda=1/3$ can be expanded with bases using $\lambda=0,\ 2/3,\ \text{and}\ 1$, as
  \begin{equation}
   \left(\cos\theta +\frac{1}{3}\right)^2 = \frac{1}{3}(\cos\theta +0)^2
    +\left(\cos\theta +\frac{2}{3}\right)^2
    -\frac{1}{3}(\cos\theta +1)^2.
    \label{eq:expansion}
  \end{equation}
  Thus one may think that the selected three bases of $\{\lambda\}=\{0.0, 0.6327, 1.0\}$ are 
  just a reducible form of the basis with $\lambda=1/3$.
  However, seeing Fig.~\ref{fig:angular-weights}, which shows the weights from the three angular bases in the NN structure,
  the three bases contribute to nodes in the hidden layer in many ways, not like Eq.~\eqref{eq:expansion}.
  Moreover, the nodes in the hidden layer receive contributions not only from angular bases but also from two-body bases,
  and the energy of an atom-$i$ is obtained from a weighted sum of these node values as in Eq.~\eqref{eq:E_i}.
  Therefore, it is difficult to know which bases are important and 
  to determine the number of bases heuristically.

\begin{figure}[tb]
   \begin{center}
    \includegraphics[keepaspectratio,width=.4\textwidth]{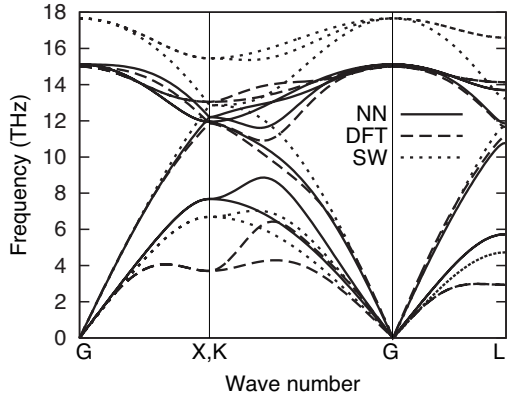}
    \caption{Phonon dispersion relations of DFT, SW, and NN.
    NN potential reproduces DFT phonon dispersion better than SW potential.}
    \label{fig:phonon}
   \end{center}
  \end{figure}

  Elastic properties of Si diamond crystal obtained using the DFT, present NN potential, 
  and SW potential are listed in Table \ref{tab:elastic-properties}.
  The NN potential reproduces elastic properties mostly within 10\%;
  the lattice constant $a_0$, bulk modulus $B$, and a component of elastic constants $C_{11}$ are in good agreement.
  However, $C_{12}$ and $C_{44}$ differ slightly from the reference values,
  since the corresponding deformation modes for them are not included in the sample set.
  If we add some structures relevant to deformation modes to the training set,
  the potential could be improved.
  Figure \ref{fig:phonon} shows the phonon dispersion relations of the NN potential, DFT, and SW potential.
  The NN potential reproduces the phonon dispersion of DFT mostly better than SW potential.
  However, transversal acoustic branches of NN at the Brillouin zone edge are not good agreement with those of DFT.
  This can also be attributed to the lack of relevant structures in sample data set,
  and the NN potential cannot well extrapolate energies or forces of such deformation modes.

  \begin{table}[tb]
   \begin{center}
    \caption{Elastic properties of diamond Si, such as lattice constant $a_0$, cohesive energy $E_{\rm coh}$,
    bulk modulus $B$, elastic constants $C_{11}$, $C_{12}$, and $C_{44}$ calculated using DFT, SW potential, and NN potential.}
    \label{tab:elastic-properties}
    \begin{tabular}[t]{lrrr} \hline\hline
     {} & DFT & SW & NN\\ \hline
     $a_0$ (${\rm \AA}$)  & 5.474 & 5.431 & 5.476 \\
     $E_{\rm coh}$ (eV/atom)  & -4.541 & -4.336 & -4.537 \\
     $B$ (GPa) & 88.31  & 101.38 & 84.08 \\
     $C_{11}$ (GPa) & 168.0 & 149.8 & 165.1 \\
     $C_{12}$ (GPa) & 77.4  & 74.8  &  64.4 \\
     $C_{44}$ (GPa) & 95.4  & 109.7 & 86.3 \\ \hline\hline
    \end{tabular}
   \end{center}
  \end{table}
  
  \begin{figure}[tb]
   \begin{center}
    \includegraphics[keepaspectratio,width=.45\textwidth]{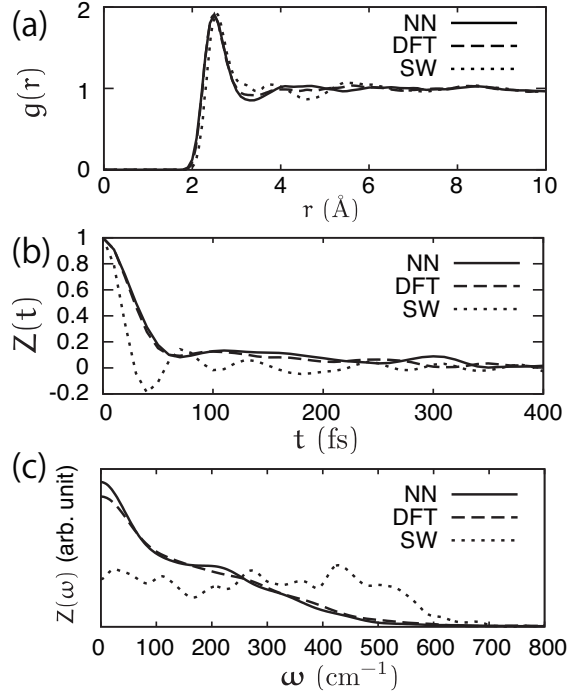}
    \caption{(a) Radial distribution function $g(r)$, (b) velocity autocorrelation function $Z(t)$, 
    and (c) its power spectrum $Z(\omega)$, of liquid-Si at 2,000 K.
    NN well reproduces DFT results compared to SW. }
    \label{fig:liqiud}
   \end{center}
  \end{figure}

  The static and dynamic properties of liquid Si are also examined and compared between the NN, DFT, and SW,
  since it is known that the SW and Tersoff potentials
  do not well reproduce the dynamic properties of liquid Si obtained by the DFT.\cite{Ishimaru:1996us,Stich:1996cv}
  Liquid structures and properties are obtained through MD simulation runs at 2,000K with about 7\% larger density
  than that of the equilibrium diamond crystal at 0K.\cite{Stich:1996cv}
  Figures \ref{fig:liqiud}(a), \ref{fig:liqiud}(b), and \ref{fig:liqiud}(c) show the radial distribution functions $g(r)$,
  velocity autocorrelation functions $Z(t)$, and their power spectra $Z(\omega)$, respectively.
  In all cases, the NN potential reproduces DFT results well, whereas the SW potential does not especially in $Z(t)$ and $Z(\omega)$.
  Especially, on the self diffusion, which can be estimated from the $Z(\omega=0)$, 
  the NN potential well reproduces DFT result much better than the SW.
  This dynamic property of liquid Si is not well reproduced if the force data are not included 
  in the function to be minimized unlike in Eq.~\eqref{eq:L}.
  Hence the force-matching is necessary when we create potentials for the purpose of dynamics simulation.

 \section{Conclusion}\label{sec:conclusion}
 
 We have constructed the neural-network (NN) interatomic potential for Si system 
 with 18 bases selected from original 300 bases through the forward stepwise (FS) regression.
 The FS regression technique, which is similar to the sparsification by the least absolute shrinkage and selection operator (LASSO),
 allows us to extract important bases for the system considered.
 This approach can be straightforwardly applicable to any compound systems and enable us to create accurate and fast NN potentials.
 The present NN potential shows good agreements with DFT results such as elastic properties and phonon dispersion relation of crystalline Si,
 and static and dynamic properties of liquid Si.
 Thus it can be used for a variety of situations such as melting, amorphization, crystallization, and mechanical deformation of Si crystal.

 \section*{Acknowledgement}
 This work was partially supported by MEXT Kakenhi 26106513 and 
 by MI$^2$I (Materials research by Information Integration Initiative) of Japan Science and Technology Agency (JST).
 

\end{document}